\documentclass[prb,aps,pre,english,superscriptaddress,floatfix]{revtex4}
\usepackage{psfig,graphics,graphicx}
\usepackage{babel}

\begin{document}

\title{Quantum-Classical Limit of Quantum Correlation Functions}

\author{Alessandro Sergi}
\email[]{asergi@chem.utoronto.ca}
\affiliation{Chemical Physics
Theory Group, Department of Chemistry, University of Toronto,
Toronto, ON M5S 3H6, Canada}
\author{Raymond Kapral}
\email[]{rkapral@chem.utoronto.ca} \affiliation{Chemical Physics
Theory Group, Department of Chemistry, University of Toronto,
Toronto, ON M5S 3H6, Canada} \affiliation{Fritz-Haber-Institut der
Max-Planck-Gesellschaft, Faradayweg 4-6, 14195 Berlin, Germany}

\date{\today}

\begin{abstract}
A quantum-classical limit of the canonical equilibrium time
correlation function for a quantum system is derived. The
quantum-classical limit for the dynamics is obtained for quantum
systems comprising a subsystem of light particles in a bath of
heavy quantum particles. In this limit the time evolution of
operators is determined by a quantum-classical Liouville operator
but the full equilibrium canonical statistical description of the
initial condition is retained. The quantum-classical correlation
function expressions derived here provide a way to simulate the
transport properties of quantum systems using quantum-classical
surface-hopping dynamics combined with sampling schemes for
the quantum equilibrium
structure of both the subsystem of interest and its environment.
\end{abstract}

\maketitle


\section{Introduction}
The dynamical properties of systems close to equilibrium may be
described in terms of equilibrium time correlation functions of
dynamical variables or operators. For a quantum system with
Hamiltonian $\hat{H}$ at temperature $T$ with volume $V$, linear
response theory shows that the time correlation function of two
operators $\hat{A}$ and $\hat{B}$, needed to obtain transport
properties, has the Kubo transformed form \cite{kubo,mori},
\begin{eqnarray}
C_{AB}(t;\beta)&=& \frac{1}{\beta}\int_0^{\beta} d \lambda {\rm
Tr} \hat{A}(t) e^{\lambda \hat{H}} \hat{B}^{\dagger}e^{-\lambda
\hat{H}} \hat{\rho}_e \nonumber \\
&=& \frac{1}{\beta}\int_0^{\beta}d\lambda \frac{1}{Z_Q}{\rm Tr}
\hat{B}^{\dag} e^{\frac{i}{\hbar}t_1^*\hat{H}}
\hat{A}e^{-\frac{i}{\hbar}t_2\hat {H}}\;, \label{eq:corfunct}
\end{eqnarray}
where $\beta=(k_B T)^{-1}$, $\hat{\rho}_e=Z_Q^{-1}e^{-\beta
\hat{H}}$ is the quantum canonical equilibrium density operator,
$Z_Q={\rm Tr} e^{-\beta \hat{H}}$ is the canonical partition
function and, in the second line, $t_1=t-i\hbar(\beta-\lambda)$
and $ t_2=t-i\hbar\lambda$. The evolution of the operator
$\hat{A}(t)$ is given by the solution of the Heisenberg equation
of motion, $d \hat{A}(t)/dt= \frac{i}{\hbar}[\hat{H},\hat{A}(t)]$,
where the square brackets denote the commutator.

While such correlation functions provide information on the
transport properties of the system, their direct computation for
condensed phase systems is not feasible due to our inability to
simulate the quantum mechanical evolution equations for systems
with a large number of degrees of freedom. While approximate
schemes have been devised to treat quantum many body dynamics, for
example, quantum mode coupling methods have proved useful in the
calculation of collective modes for some applications
\cite{rabani}, we are primarily concerned with methods that
approximate the full many body evolution of the microscopic
degrees of freedom. In many circumstances only a few degrees of
freedom need to be treated quantum mechanically (quantum
subsystem) while the remainder of the system with which they
interact can be treated classically (classical bath) to a good
approximation. Examples where such a description is appropriate
include proton and electron transfer processes occurring in
solvents or other chemical environments composed of heavy atoms.
Quantum-classical methods have been reviewed by Egorov et al.
\cite{egorov1} and one form of a quantum-classical approximation
has been assessed in the weak coupling limit where there is no
feedback between the quantum and classical subsystems. Although it
is difficult to determine transport properties such as the
reaction rate constant from the full quantum time correlation
function when the entire system is treated quantum mechanically,
methods are being developed to carry out such calculations.
\cite{golosov} Mixed quantum-classical methods also provide a
route to carry out nonadiabatic rate calculations.

A number of schemes have been proposed for carrying out quantum
dynamics in classical environments.
\cite{pechukas,herman,tullyrev,tully,coker,rossky,martinez,warshel}
We focus on approaches where the evolution is described by a
quantum-classical Liouville equation.
\cite{alek,geras,boucher,balescu,martens,schutte,schofield,kapral:1}
For a quantum system coupled to a classical environment it is
possible to derive an evolution equation for dynamical variables or
operators (or the density matrix) by an expansion in a small parameter
that characterizes the mass ratio of the light and heavy particles in
the system. The quantum-classical analog of the Heisenberg equation
of motion is, \cite{kapral:1}
\begin{eqnarray}
\frac{d } {d t}\hat{A}_W(X,t)&=&\frac{i}{\hbar}
[\hat{H}_W, \hat{A}_W(t)]
-\frac{1}{2} \Big( \{\hat{H}_W,\hat{A}_W(t)\}
- \{\hat{A}_W(t), \hat{H}_W\} \Big)\nonumber \\
&=& i \hat{\mathcal L}\hat{A}_W(X,t) \;. \label{eq:dmatabs}
\end{eqnarray}
Here $\hat{A}_W(X,t)$ is the partial Wigner representation
\cite{wigner,kapral:1} of a quantum operator; it is still an
operator in the Hilbert space of the quantum subsystem but a
function of the phase space coordinates $X=(R,P)$ of the classical
bath.  In this equation $\{\cdots,\cdots\}$ is the Poisson bracket
and $\hat{\mathcal L}$ is the quantum-classical Liouville
operator. A few features of quantum-classical Liouville dynamics
are worth noting. This equation of motion includes feedback
between the classical and quantum degrees of freedom. The
environmental dynamics is fully classical only in the absence of
coupling to the quantum subsystem. In the presence of coupling the
environmental evolution cannot be described by Newtonian dynamics,
although the simulation of the quantum-classical evolution can be
formulated in terms of classical trajectory segments.
\cite{kapral:1} For harmonic environmental potentials with
bilinear coupling to the quantum subsystem the evolution is
equivalent to the fully quantum mechanical evolution of the entire
system. Quantum-classical simulations of the spin-boson model are
in accord with the numerically exact quantum results \cite{makri}
and have been used to test quantum-classical simulation
algorithms. \cite{donal:1,slice}

Equation~(\ref{eq:dmatabs}) is valid in any basis and an
especially convenient basis for simulating the evolution by
surface-hopping schemes is the adiabatic basis, $\{|\alpha_1;
R>\}$, the set of eigenstates of the quantum subsystem Hamiltonian
in the presence of fixed classical particles. In this case the
matrix elements of an operator $A_W^{\alpha_1
\alpha_1'}(X,t)=<\alpha_1;R|\hat{A}_W(X,t)|\alpha_1';R>$ satisfy
\begin{equation}
\frac{d}{dt}A_W^{\alpha_1 \alpha_1'}(X,t)= i\sum_{\beta_1
\beta_1'} {\mathcal L}_{\alpha_1 \alpha_1' \beta_1 \beta_1'}
A_W^{\beta_1 \beta_1'}(X,t)\;, \label{eq:qcamat}
\end{equation}
where ${\mathcal L}_{\alpha_1 \alpha_1' \beta_1 \beta_1'}$ denotes
the representation of the quantum-classical Liouville operator in
the adiabatic basis. \cite{kapral:1} This equation may be solved
using surface-hopping schemes that combine a probabilistic
description of the quantum transitions interspersed with classical
evolution trajectory segments.
\cite{steve:1,steve:2,donal:1,slice,review} Although further
algorithm development is needed to carry the simulations to
arbitrarily long times, the quantum-classical evolution is not a
short time approximation to full quantum evolution. Rather, it is
an approximation to the full quantum evolution for arbitrary times
since the quantum-classical evolution is derived at the level of
the Liouville operator and not the quantum propagator. Given the
evolution equation (\ref{eq:dmatabs}) (and the corresponding
quantum-classical Liouville equation for the density matrix) one
may construct a statistical mechanics of quantum-classical systems
\cite{mqcstat,simu} and compute transport properties such as
chemical rate constants \cite{mqcrate} from the correlation
functions obtained from this analysis.

In this article we consider another route to determine
quantum-classical correlation functions for transport properties.
We begin with the full quantum mechanical expression for the time
correlation function (Eq.~(\ref{eq:corfunct})) and take the
limit where the dynamics is determined by quantum-classical
evolution equations for the spectral density that enters the
correlation function expression. While the calculations leading to
our expression for the correlation function are somewhat lengthy,
the final result has a simple structure:
\begin{eqnarray}
C_{AB}(t;\beta) &=& \sum_{\beta_1'\beta_1\beta_2'\beta_2} \int
dX_1 dX_2 \; B_W^{\dag
\beta_1\beta_1'}(X_1,\frac{t}{2})\nonumber \\
&&\times A_W^{\beta_2\beta_2'}(X_2,-\frac{t}{2})
\overline{W}^{\beta_1'\beta_1\beta_2'\beta_2}(X_1,X_2;\beta)
\;.\nonumber
\\ \label{eq:Iqccor}
\end{eqnarray}
This expression for the time correlation function retains the full
quantum statistical character of the initial condition through the
spectral density function $\overline{W}$ (Eq.~(\ref{eq:overW})
below) but the forward and backward time evolution of the
operators $B_W^{\dag \beta_1\beta_1'}$ and
$A_W^{\beta_2\beta_2'}$, respectively, is given by the solutions
of the quantum-classical evolution equation (\ref{eq:qcamat}).
Consequently, one may combine algorithms for determining quantum
equilibrium properties with surface-hopping algorithms for
quantum-classical evolution to estimate the value of the
correlation function. Quantum effects enter in all orders in this
expression for the correlation function. In addition to the fact
that the initial value of the spectral density contains the full
quantum equilibrium statistics, since the quantum-classical
Liouville operator appears in the exponent in the propagator, the
quantum-classical propagator contains all orders of $\hbar$,
albeit in an approximate fashion.

The outline of the paper is as follows: In Sec.~\ref{sec:pWW} we
construct the partial Wigner representation of the quantum time
correlation function and obtain expressions for the spectral
density and its matrix elements in an adiabatic basis. In
Sec.~\ref{sec:qcW} we derive a quantum-classical evolution
equation for the matrix elements of the spectral density and
establish a connection to quantum-classical Liouville evolution.
The results of these two sections are used in Sec.~\ref{sec:qccor}
to obtain Eq.~(\ref{eq:Iqccor}). In Sec.~\ref{sec:icondt} we
analyze the initial value of the spectral density in the high
temperature limit. The conclusions of the study are given in
Sec.~\ref{sec:conc} while additional details of the calculations
are presented in the Appendices.

\section{Partial Wigner Representation of Quantum Correlation
Function}\label{sec:pWW}

We consider quantum systems whose degrees of freedom can be
partitioned into two subsets corresponding to light (mass $m$) and
heavy (mass $M$) particles, respectively. We use small and capital
letters to denote operators for phase points in the light and heavy
mass subsystems, respectively. In this notation the Hamiltonian
operator for the entire system is the sum of the kinetic energy
operators or the two subsystems and the potential energy of the
entire system,
$\hat{H}=\hat{P}^2/2M+\hat{p}^2/2m+\hat{V}(\hat{q},\hat{Q})$.

We are interested in the limit where the dynamics of the heavy
particle subsystem is treated classically and the light particle
subsystem retains its full quantum character. To this end it is
convenient to take a partial Wigner transform of the heavy degrees
of freedom and represent the light degrees of freedom in some
suitable basis.

In order to carry out this program we begin with the quantum 
mechanical Kubo transformed correlation function and
write the trace over the
heavy subsystem degrees of freedom in the second line of Eq.~(\ref{eq:corfunct})
using a $\{Q\}$ coordinate representation,
\begin{eqnarray}
C_{AB}(t;\beta) &=& \frac{1}{\beta}\int_0^{\beta}d\lambda
\frac{1}{Z_Q}{\rm Tr}' \int \prod_{i=1}^4 dQ_i \; <Q_1|
\hat{B}^{\dag}|Q_2> <Q_2| e^{\frac{i}{\hbar}t_1^*\hat{H}}|Q_3>
<Q_3|\hat{A}|Q_4><Q_4|e^{-\frac{i}{\hbar}t_2\hat {H}}|Q_1> \;.
\label{eq:corfQrep}
\end{eqnarray}
The prime in ${\rm Tr}'$ refers to the fact that the trace is now
only over the light particle subsystem degrees of freedom. Making
use of the change of variables, $Q_1=R_1-Z_1/2$, $Q_2=R_1+Z_1/2$,
$Q_3=R_2-Z_2/2$ and $Q_4=R_2+Z_2/2$, this equation may be written
in the equivalent form,
\begin{eqnarray}
C_{AB}(t;\beta) &=& \frac{1}{\beta}\int_0^{\beta}d\lambda
\frac{1}{Z_Q}{\rm Tr}' \int \prod_{i=1}^2 dR_i dZ_i\;
<R_1-\frac{Z_1}{2}| \hat{B}^{\dag}|R_1+\frac{Z_1}{2}>
<R_1+\frac{Z_1}{2}|e^{\frac{i}{\hbar}t_1^*\hat{H}}|R_2-\frac{Z_2}{2}>
\nonumber \\
&&\times
<R_2-\frac{Z_2}{2}|\hat{A}|R_2+\frac{Z_2}{2}>
<R_2+\frac{Z_2}{2}|e^{-\frac{i}{\hbar}t_2\hat
{H}}|R_1-\frac{Z_1}{2}> \;. \label{eq:corfQ2rep}
\end{eqnarray}
The next step in the calculation is to replace the coordinate
space matrix elements of the operators with their representation
in terms of Wigner transformed quantities. The partial Wigner
transform of an operator is defined by \cite{wigner,kapral:1}
\begin{equation}
\hat{A}_W(R_2,P_2)= \int dZ_2 e^{\frac{i}{\hbar}P_2\cdot Z_2}
\langle R_2-\frac{Z_2}{2}|\hat{A}|R_2+\frac{Z_2}{2}\rangle \;,
\label{eq:pwig_t}
\end{equation}
while the inverse transform is
\begin{equation}
\langle R_2-\frac{Z_2}{2}|\hat{A}|R_2+\frac{Z_2}{2}\rangle
=\frac{1}{(2\pi\hbar)^{\nu_h}} \int dP_2 e^{-\frac{i}{\hbar}P_2
\cdot Z_2}\hat{A}_W(R_2,P_2)\;.
\label{eq:Winv}
\end{equation}
Here $\nu_h$ is the dimension of the heavy mass subsystem. The
partially Wigner transformed operator $\hat{A}_W(X_2)$ is a
function of the phase space coordinates $X_2 \equiv (R_2,P_2)$ and
an operator in the Hilbert space of the quantum subsystem. It is
convenient to consider a representation of such operators in basis
of eigenfunctions. In this paper we use an adiabatic basis since,
through this representation, we can make connection with
surface-hopping dynamics. The partial Wigner transform of the
Hamiltonian $\hat{H}$ is
$\hat{H}_W=P^2/2M+\hat{p}^2/2m+\hat{V}_W(\hat{q},R)\equiv
P^2/2M+\hat{h}_W(R)$. The last equality defines the Hamiltonian
$\hat{h}_W(R)$ for the light mass subsystem in
the presence of fixed particles of the heavy mass subsystem. The
adiabatic basis is determined from the solutions of the eigenvalue
problem, $\hat{h}_W(R)|\alpha;R>=E_{\alpha}(R)|\alpha;R>$. The
adiabatic representation of $\hat{A}_W(X_2)$ is,
\begin{equation}
\hat{A}_W(X_2)=\sum_{\alpha_2 \alpha_2'} |\alpha_2;R_2>A^{\alpha_2
\alpha_2'}_W(X_2)<\alpha_2';R_2| \;, \label{eq:Abasis}
\end{equation}
where $A^{\alpha_2
\alpha_2'}_W(X_2)=<\alpha_2;R_2|\hat{A}_W(X_2)|\alpha_2';R_2>$.

By inserting Eq.~(\ref{eq:Abasis}) into Eq.~(\ref{eq:Winv}) we can
express the coordinate representation of the operator $\hat{A}$ as
\begin{equation}
<R_2-\frac{Z_2}{2}|\hat{A}|R_2+\frac{Z_2}{2}>= \frac{1}{(2 \pi
\hbar)^{\nu_h}} \sum_{\alpha_2 \alpha_2'} \int dP_2 \;
e^{-\frac{i}{2}P_2 \cdot Z_2} |\alpha_2;R_2>A_W^{\alpha_2
\alpha_2'}(X_2)<\alpha_2';R_2| \;.
\end{equation}
Then, substituting the result into Eq.~(\ref{eq:corfQ2rep}) (along
with a similar representation of the $\hat{B}^\dag$ operator), we
obtain,
\begin{eqnarray}
C_{AB}(t;\beta) &=& \frac{1}{\beta}\int_0^{\beta}d\lambda
\sum_{\alpha_1,\alpha_1',\alpha_2,\alpha_2'} \int \prod_{i=1}^2
dX_i \;  B_W^{\dag \alpha_1\alpha_1'}(X_1)
A_W^{\alpha_2\alpha_2'}(X_2)
\nonumber\\
&\times&
W^{\alpha_1'\alpha_1\alpha_2'\alpha_2}(X_1,X_2,t;\beta,\lambda)\;.
\label{eq:corfin1}
\end{eqnarray}
Here we defined the matrix elements of the spectral density by
\begin{eqnarray}
W^{\alpha_1'\alpha_1\alpha_2'\alpha_2}(X_1,X_2,t;\beta,\lambda)&=&
\frac{1}{Z_Q} \int \prod_{i=1}^2 dZ_i e^{-\frac{i}{\hbar}(P_1\cdot
Z_1+P_2\cdot Z_2)}
<\alpha_1';R_1|<R_1+\frac{Z_1}{2}|e^{\frac{i}{\hbar}t_1^*\hat{H}}
|R_2-\frac{Z_2}{2}>|\alpha_2;R_2>
\nonumber \\
&&\times
<\alpha_2';R_2|<R_2+\frac{Z_2}{2}|e^{-\frac{i}{\hbar}t_2\hat
{H}}|R_1-\frac{Z_1}{2}>|\alpha_1;R_1>\frac{1}{(2\pi\hbar)^{2\nu_h}}\;.
\label{eq:Wmatel}
\end{eqnarray}
Our task is now to find an evolution equation for
$W^{\alpha_1'\alpha_1\alpha_2'\alpha_2}(X_1,X_2,t;\beta,\lambda)$
in the mixed quantum-classical limit. Before doing this we observe
that the expression for the quantum correlation function in the
partial Wigner representation is equivalent to an expression
involving full Wigner transforms of the operators.

\subsection{Relation to full Wigner representation}
Since the correlation function is independent of the
representation we choose for the light and heavy mass subsystems,
we may also represent the light mass subsystem in terms of a
Wigner transform instead of a set of basis functions. To establish
the connection between these two forms of the correlation function
we note that the full Wigner transform of the the operator
$\hat{A}$ is given by
\begin{eqnarray}
A_W(x_2,X_2)&=& \int dz_2\; e^{\frac{i}{\hbar}p_2\cdot z_2}
<r_2-\frac{z_2}{2}|\hat{A}_W(X_2)|r_2+\frac{z_2}{2}>\;, \nonumber
\\
&=&\sum_{\alpha_2 \alpha_2'}\int dz_2\; e^{\frac{i}{\hbar}p_2\cdot
z_2} \phi_{\alpha_2}(r_2-\frac{z_2}{2};R_2)A^{\alpha_2
\alpha_2'}_W(X_2)\phi_{\alpha_2'^*}(r_2+\frac{z_2}{2};R_2)\;,
\end{eqnarray}
where $x_2=(r_2,p_2)$ and
$\phi_{\alpha_2}(r_2;R_2)=<r_2|\alpha_2;R_2>$. We have used
Eq.~(\ref{eq:Abasis}) to write the second line of this equation.
The inverse of this expression is
\begin{eqnarray}
A^{\alpha_2 \alpha_2'}_W(X_2)=
\frac{1}{(2\pi\hbar)^{\nu_{\ell}}}\int dp_2
d(r_2-\frac{z_2}{2})d(r_2+\frac{z_2}{2})\;
e^{-\frac{i}{\hbar}p_2\cdot z_2}
A_W(x_2,X_2)\phi_{\alpha_2}^*(r_2-\frac{z_2}{2};R_2)
\phi_{\alpha_2'}(r_2+\frac{z_2}{2};R_2)\;,
\end{eqnarray}
where $\nu_{\ell}$ is the dimension of the light mass subsystem.
Inserting this expression for $A^{\alpha_2 \alpha_2'}_W(X_2)$ (and
the analogous expression for $B_W^{\dag\alpha_1\alpha_1'}(X_1)$)
into Eq.~(\ref{eq:corfin1}) we find
\begin{eqnarray}
C_{AB}(t,\beta)&=&\frac{1}{\beta} \int_0^{\beta}d\lambda\int
\prod_{i=1}^2 dx_i dX_i\;
B_W^{\dag}(x_1,X_1)A_W(x_2,X_2)\nonumber\\
&&\times W(x_1,X_1,x_2,X_2;t,\beta,\lambda) \;,
\label{eq:fullwig-corr}
\end{eqnarray}
where
\begin{eqnarray}
W(x_1,x_2,X_1,X_2,t;\beta,\lambda)
&=&\sum_{\alpha_1'\alpha_1\alpha_2'\alpha_2} \int \prod_{i=1}^2
dz_i\; e^{-\frac{i}{\hbar}(p_1\cdot z_1+p_2 \cdot z_2)}
\phi_{\alpha_1'}(r_1+\frac{z_1}{2};R_1)\phi_{\alpha_1}^*(r_1-\frac{z_1}{2};R_1)
\nonumber \\
&&\times \phi_{\alpha_2'}(r_2+\frac{z_2}{2};R_2)
\phi_{\alpha_2}^*(r_2-\frac{z_2}{2};R_2)
W^{\alpha_1'\alpha_1
\alpha_2'\alpha_2}(X_1,X_2,t;\beta,\lambda)\frac{1}{(2\pi\hbar)^{2\nu_l}}\;.
\label{eq:WFpartWa}
\end{eqnarray}
Using the definition of the matrix elements of $W$ in
Eq.~(\ref{eq:Wmatel}) and performing the sums on states we may
write this equation in the equivalent form,
\begin{eqnarray}
W(x_1,x_2,X_1,X_2,t;\beta,\lambda)&=& \frac{1}{Z_Q}\int
\prod_{i=1}^2 dz_i dZ_i \; e^{-\frac{i}{\hbar}(p_1\cdot
z_1+p_2\cdot z_2)} e^{-\frac{i}{\hbar}(P_1\cdot Z_1+P_2\cdot Z_2)}\nonumber \\
&&\times
<r_1+\frac{z_1}{2}|<R_1+\frac{Z_1}{2}|e^{\frac{i}{\hbar}t_1^*\hat{H}}
|R_2-\frac{Z_2}{2}>|r_2-\frac{z_2}{2}>
\nonumber \\
&&\times
<r_2+\frac{z_2}{2}|<R_2+\frac{Z_2}{2}|e^{-\frac{i}{\hbar}t_2\hat
{H}}|R_1-\frac{Z_1}{2}>|r_1-\frac{z_1}{2}>\frac{1}{(2\pi\hbar)^{2\nu}}\;,
\label{eq:WFpartW}
\end{eqnarray}
where $\nu=\nu_{\ell}+\nu_h$. Equation~(\ref{eq:WFpartW}) gives
the spectral density in the full Wigner representation while
Eq.~(\ref{eq:WFpartWa}) relates this quantity to its matrix
elements in the light mass subsystem basis. In particular, letting
${\bf W}$ be a (super) matrix whose elements are
$W^{\alpha_1'\alpha_1 \alpha_2'\alpha_2}$, Eq.~(\ref{eq:WFpartWa})
may be written formally as
\begin{equation}
W= \mbox{\boldmath${\mathcal T}$} \circ {\bf W}\;, \label{eq:Tdef}
\end{equation}
where $\mbox{\boldmath${\mathcal T}$}$ is the transformation
specified by Eq.~(\ref{eq:WFpartWa}).

For future use, we note that the inverse of this expression is
given by
\begin{eqnarray}
W^{\alpha_1'\alpha_1\alpha_2'\alpha_2}(X_1,X_2,t; \beta,\lambda)
&=& \int \prod_{i=1}^2 dx_i dz_i \; e^{\frac{i}{\hbar}(p_1 \cdot
z_1+p_2 \cdot z_2)}
\phi_{\alpha_1}(r_1-\frac{z_1}{2};R_1)\phi_{\alpha_1'}^*(r_1+\frac{z_1}{2};R_1)
\nonumber\\
&&\times
\phi_{\alpha_2}(r_2-\frac{z_2}{2};R_2)\phi_{\alpha_2'}^*(r_2+\frac{z_2}{2};R_2)
W(x_1,x_2,X_1,X_2,t; \beta,\lambda)\;, \label{eq:partWFW}
\end{eqnarray}
as can be verified by substituting Eq.~(\ref{eq:WFpartW}) into
Eq.~(\ref{eq:partWFW}) and performing the integrals. Like
Eq.~(\ref{eq:WFpartWa}), Eq.~(\ref{eq:partWFW}) gives a relation
between $W$ and its matrix elements but now in the opposite
direction, relating $W^{\alpha_1'\alpha_1\alpha_2'\alpha_2}$ to
$W$. Using a formal notation like that in Eq.~(\ref{eq:Tdef}), we
can write Eq.~(\ref{eq:partWFW}) as
\begin{equation}
{\bf W}= \mbox{\boldmath${\mathcal T}$}^{-1} \circ W \;,
\label{eq:Tinvdef}
\end{equation}
which defines $\mbox{\boldmath${\mathcal T}$}^{-1}$, the inverse
of $\mbox{\boldmath${\mathcal T}$}$.

\section{Quantum-Classical Evolution Equation for Spectral
Density} \label{sec:qcW}

The quantum-classical evolution equation for matrix elements of
the spectral density
$W^{\alpha_1'\alpha_1\alpha_2'\alpha_2}(X_1,X_2,t;\beta,\lambda)$
can be obtained using various routes. In this paper we first
derive a quantum-classical evolution equation for the spectral
density $W(x_1,x_2,X_1,X_2,t;\beta,\lambda)$ in the full Wigner
representation. We then change to an adiabatic basis
representation of the quantum subsystem to obtain our final result
for the evolution equation for
$W^{\alpha_1'\alpha_1\alpha_2'\alpha_2}(X_1,X_2,t;\beta,\lambda)$.

\subsection{Evolution of $W$ in the full Wigner representation}

The evolution equation for $W(x_1,x_2,X_1,X_2,t;\beta,\lambda)$
can be obtained by differentiating its definition in
Eq.~(\ref{eq:WFpartW}) with respect to time and then
inserting complete sets of coordinate states to obtain a closed
equation in $W$. The result was obtained earlier by Filinov et al.
\cite{filinov} and, for our composite system, is given by
\begin{eqnarray}
\frac{\partial }{\partial t}W(x_1,x_2,X_1,X_2,t;\beta,\lambda)&=&
-\frac{1}{2}\Big(iL^{(0)}_1(x_1,X_1)-iL^{(0)}_2(x_2,X_2)\Big)
W(x_1,x_2,X_1,X_2,t;\beta,\lambda)
\nonumber \\
&+&\frac{1}{2}\int \prod_{i=1}^2 ds_i dS_i \;
\Big(\omega_1(r_1,s_1,R_1,S_1)\delta(s_2)\delta(S_2) -
\omega_2(r_2,s_2,R_2,S_2)\delta(s_1)\delta(S_1)\Big)
\nonumber \\
&\times&
W(x_1-\pi_1,X_1-\Pi_1,x_2-\pi_2,X_2-\Pi_2,t;\beta,\lambda)
\;.\label{eq:partW}
\end{eqnarray}
Here we have introduced the notation $\pi_i=(0,s_i)$,
$\Pi_i=(0,S_i)$. The classical free streaming Liouville operators
are
\begin{eqnarray}
iL^{(0)}_i(x_i,X_i)= \frac{p_i}{m}\cdot \frac{\partial}{\partial
q_i} +\frac{P_i}{M}\cdot \frac{\partial}{\partial R_i} \;,
\end{eqnarray}
for $(i=1,2)$. The $\omega_i$ functions under the integral are
defined by
\begin{eqnarray}
\omega_i(r_i,R_i,s_i,S_i)&=&\frac{2}{\hbar(\pi\hbar)^{\nu}}
\int d\bar{r}_id\bar{R}_i \; V(r_i-\bar{r}_i,R_i-\bar{R}_i)\nonumber\\
&\times& \sin(\frac{2}{\hbar} s_i \cdot \bar{r}_i+\frac{2}{\hbar}
S_i \cdot \bar{R}_i) \;. \label{eq:omega}
\end{eqnarray}

\subsection{Scaled equation of motion}

In order to take the quantum-classical limit of
Eq.~(\ref{eq:partW}), we consider systems for which the ratio
between the light particle mass and the heavy particle mass is
small, and employ the same mass scaling used in
Ref.~\cite{kapral:1} to obtain the quantum-classical Liouville
equation. One is naturally led to consider an expansion in the
small parameter $\mu=(m/M)^{1/2}$ from the following arguments.
Consider a unit of energy $\epsilon_0$, say the thermal energy
$\epsilon_0=\beta^{-1}$, a unit of length $\lambda_m=\hbar/p_m$,
where $p_m=(m\epsilon_0)^{1/2}$ is the unit of momentum of the
light particles, a unit of time $t_0=\hbar\epsilon_0^{-1}$ and a
unit of momentum for the heavy particles
$P_M=(M\epsilon_0)^{1/2}$. These units may be used to scale the
coordinates of the system so that the magnitude of the scaled
momentum of the heavy particles, $P/P_M$, is of the same order of
magnitude as that for the light particles, $p/p_m$. Only momenta
are scaled by different factors; characteristic lengths are scaled
by the light particle thermal de Broglie wavelength, $\lambda_m$.
This is analogous to the scaling used to derive the equations of
Brownian motion for a heavy particle in a bath of light particles.

In the following, we denote scaled quantities with a prime; e.g.,
$r'=r/\lambda_m$, $R'=R/\lambda_m$, $p'=p/p_m$, $P'=P/P_M$,
$t'=t/t_0$, etc. The scaled version of the equation of motion for
$W$ has the same form as Eq.~(\ref{eq:partW}) but with all
quantities replaced by their primed dimensionless counterparts.
The scaled operators and functions in this equation are
\begin{eqnarray}
iL^{(0)\prime}_i(x_i',X_i')= p_i'\cdot \frac{\partial}{\partial
q_i'} +\mu P_i'\cdot \frac{\partial}{\partial R_i'} \;,
\label{eq:scaledL0}
\end{eqnarray}
and
\begin{eqnarray}
\omega_i'(r_i',R_i',s_i',S_i')&=& \frac{2}{\pi^{\nu}}
\int d\bar{r}_i'd\tilde{R}_i' \; V'(r_i'-\bar{r}_i',R_i'-\mu\tilde{R}_i')\nonumber\\
&\times& \sin(2 s_i'\cdot \bar{r}_i'+2 S_i' \cdot \tilde{R}_i')
\;. \label{eq:scaledomega}
\end{eqnarray}
In writing the last line of the $\omega'_i$ equation we have
performed the change of variables $\tilde{R}_1=\mu^{-1}\bar{R}_1$
in the dummy variable in the integration in order to move the
$\mu$ dependence from the sine factor to the potential, which is
more convenient for taking the classical limit.
We see that
the classical free streaming evolution is linear in $\mu$ but the
quantum kernel has all powers of $\mu$.

\subsection{Quantum-classical equation for $W$}

For $\mu<<1$ the quantum-classical limit
is obtained expanding the evolution operator up to linear terms in
$\mu$. Since momentum is related to the de Broglie wavelength of a
particle, this procedure is equivalent to averaging out the short
de Broglie oscillations of the heavy particles on the scale of the
long de Broglie oscillations of the light particles.

The expansion of the evolution operator is obtained from the
expansion of the interaction potential to linear order in the
small parameter $\mu$,
\begin{equation}
V'(r_i'-\bar{r}_i',R_i'-\mu \tilde{R}_i')=V(r_i'-\bar{r}_i',R_i')
-\mu\tilde{R}_i' \cdot \frac{\partial
V'(r_i'-\bar{r}_i',R_i')}{\partial R_i'}+{\cal O}(\mu^2) \;.
\label{eq:potexp}
\end{equation}
Inserting this expansion in Eq.~(\ref{eq:scaledomega}), working
out the integrals, substituting the result into the scaled version
of  Eq.~(\ref{eq:partW}) and finally going back to unscaled
coordinates (the details are given in Appendix~\ref{app:qcW}) one
obtains the quantum-classical equation for $W$ in unscaled
coordinates as
\begin{eqnarray}
&&\frac{\partial}{\partial t} W(x_1,x_2,X_1,X_2,t;\beta,\lambda)
=-\frac{1}{2}\left[iL^{(0)}_1(x_1)+iL_1(X_1)-iL^{(0)}_2(x_2)-iL_2(X_2)\right]
W(x_1,x_2,X_1,X_2,t;\beta,\lambda)
\nonumber\\
&&\quad +\int ds_1ds_2 \left\{
\frac{1}{\hbar(\pi\hbar)^{\nu_{\ell}}} \int d\bar{r} \left[
\delta(s_2)V(r_1-\bar{r},R_1)\sin\left(\frac{2s_1 \cdot
\bar{r}}{\hbar}\right)
-\delta(s_1)V(r_2-\bar{r},R_2)\sin\left(\frac{2s_2 \cdot
\bar{r}}{\hbar} \right)\right] \right.
\nonumber\\
&&\left.\quad +\frac{1}{2}\left[ \delta(s_2)\Delta
F_1(R_1,s_1)\cdot \frac{\partial}{\partial P_1} -\delta(s_1)
\Delta F_2(R_2,s_2)\cdot \frac{\partial}{\partial
P_2}\right]\right\} W(x_1-\pi_1,x_2-\pi_2,X_1,X_2,t;\beta,\lambda)
\;. \label{eq:qcwignerW}
\end{eqnarray}
Here we have defined full classical Liouville operator for the
heavy mass subsystem as,
\begin{equation}
iL_i(X_i)= \frac{P_i}{M}\cdot \frac{\partial}{\partial R_i} +
F_{R_i}\cdot \frac{\partial}{\partial P_i}\;,
\end{equation}
for $(i=1,2)$, where the force $F_{R_i}=-\partial
V(r_i,R_i)/\partial R_i$. We have also introduced the quantity
\begin{eqnarray}
\Delta F_i(R_i,s_i)&=&-\frac{\partial}{\partial
R_i}\left(V(r_i,R_i)\delta(s_i)
-\frac{1}{(\pi\hbar)^{\nu_{\ell}}}\int d\bar{r}
V(r_i-\bar{r},R_i)\cos(\frac{2s_i \bar{r}}{\hbar})\right) \;.
\label{eq:DeltaF_i}
\end{eqnarray}
If the potential is decomposed into light and heavy mass subsystem
potentials, $V_{\ell}(r_i)$ and $V_h(R_i)$, respectively, and
their interaction potential $V_c(r_i,R_i)$ as
$V=V_{\ell}+V_h+V_c$, it is easy to demonstrate that $\Delta
F_i(R_i,s_i)$ depends only on the interaction potential.

The quantum-classical evolution equation (\ref{eq:qcwignerW}) for
$W$ can be written formally and compactly as
\begin{equation}
\frac{\partial}{\partial t}W(t)= \frac{1}{2}K \circ W(t)\;,
\label{eq:fqcwignerW}
\end{equation}
where the operator $K$ is defined by comparison with
Eq.~(\ref{eq:qcwignerW}). To simplify the notation, we have
dropped the classical phase space arguments here and some of the
following equations when confusion is unlikely to arise.

This quantum-classical evolution equation for the spectral density
is not yet in a convenient form for simulation since the kernels
that appear in this equation are highly oscillatory functions
arising from the fact that a Wigner representation of the quantum
degrees of freedom has been used. In the next subsection we
re-introduce the adiabatic basis and obtain a form of the
quantum-classical evolution equation that can be solved by
surface-hopping schemes.

\subsection{Quantum-classical evolution equation for
$W^{\alpha_1'\alpha_1\alpha_2'\alpha_2}$}

The operators $\mbox{\boldmath${\mathcal T}$}$ and
$\mbox{\boldmath${\mathcal T}$}^{-1}$ can be used to convert
Eq.~(\ref{eq:fqcwignerW}) into an evolution equation for ${\bf
W}$, the matrix elements of $W$. Acting from the left with
$\mbox{\boldmath${\mathcal T}$}^{-1}$ on Eq.~(\ref{eq:fqcwignerW})
and inserting unity in the form $W=\mbox{\boldmath${\mathcal T}$}
\circ \mbox{\boldmath${\mathcal T}$}^{-1} \circ W
=\mbox{\boldmath${\mathcal T}$} \circ {\bf W}$, we obtain,
\begin{eqnarray}
\frac{\partial}{\partial t}{\bf W}(t)&=&
\frac{1}{2}(\mbox{\boldmath${\mathcal T}$}^{-1} \circ K \circ
\mbox{\boldmath${\mathcal T}$}) \circ {\bf W}(t)\nonumber \\
&\equiv& \frac{1}{2} \mbox{\boldmath ${\mathcal K}$} \circ {\bf
W}(t)\;,\label{eq:fqcwignerWmat}\;.
\end{eqnarray}
The transformed operator on right hand side of
Eq.~(\ref{eq:fqcwignerWmat}) can be calculated explicitly by a
straightforward but lengthy calculation which is given in detail
in Appendix~\ref{t-1utw}. The result of this calculation is
\begin{eqnarray}
\frac{\partial } {\partial t}
W^{\alpha_1'\alpha_1\alpha_2'\alpha_2}(X_1,X_2,t;\beta,\lambda)
&=&\frac{1}{2}\sum_{\beta_1'\beta_1\beta_2'\beta_2} {\mathcal
K}_{\alpha_1'\alpha_1\alpha_2'\alpha_2,\beta_1'\beta_1\beta_2'\beta_2}
W^{\beta_1'\beta_1\beta_2'\beta_2}(X_1,X_2,t;\beta,\lambda)
\nonumber\\
&=&\frac{1}{2}\sum_{\beta_1'\beta_1\beta_2'\beta_2} \Big(-i{\cal
L}_{\alpha_1'\alpha_1,\beta_1'\beta_1}(X_1)
\delta_{\alpha_2'\beta_2'}\delta_{\alpha_2\beta_2} +i{\cal
L}_{\alpha_2'\alpha_2,\beta_2'\beta_2}(X_2)
\delta_{\alpha_1'\beta_1'}\delta_{\alpha_1\beta_1}\Big) \nonumber
\\
&&\times
W^{\beta_1'\beta_1\beta_2'\beta_2}(X_1,X_2,t;\beta,\lambda)
\;.\label{eq:eqofm}
\end{eqnarray}

We see that the apparently formidable evolution operator
$\mbox{\boldmath ${\mathcal K}$}$ for the matrix elements of the
spectral density, which depends on eight quantum indices, takes a
simple form consisting of a difference of two quantum-classical
Liouville operators, each acting separately on the classical phase
space variables and quantum indices with labels 1 and 2. The
quantum-classical Liouville operators are just those obtained in
earlier derivations of the quantum-classical Liouville equation
\cite{kapral:1},
\begin{eqnarray}
i{\cal L}_{\alpha_i'\alpha_i,\beta_i'\beta_i}(X_i)
=\Big(i\omega_{\alpha_i'\alpha_i}(R_i)+iL_{\alpha_i'\alpha_i}(X_i)\Big)
\delta_{\alpha_i'\beta_i'}\delta_{\alpha_i\beta_i}
-J_{\alpha_i'\alpha_i,\beta_i'\beta_i}(X_i)\;,
\end{eqnarray}
where $\omega_{\alpha
\alpha'}(R)=(E_{\alpha}(R)-E_{\alpha'}(R))/\hbar$ and
\begin{eqnarray}
iL_{\alpha_i'\alpha_i}(X_i)&=&\frac{P_i}{M}\cdot
\frac{\partial}{\partial R_i}
+\frac{1}{2}\left(F_W^{\alpha_i'}(R_i)+F_W^{\alpha_i}(R_i)\right)
\cdot \frac{\partial}{\partial P_i}\;,
\end{eqnarray}
is the classical Liouville operator involving the mean of the
Hellmann-Feynman forces where $F_W^{\alpha}= - \langle \alpha; R |
{\partial \hat{V}_W(\hat{q},R)\over \partial R}|\alpha;R \rangle
=-\langle \alpha; R | {\partial \hat{H}_W(R)\over \partial
R}|\alpha;R \rangle $. Quantum transitions and bath momentum
changes are described by
\begin{eqnarray}
J_{\alpha_i'\alpha_i,\beta_i'\beta_i}(X_i) &=& -\frac{P_i}{M}
\cdot d_{\alpha_i'\beta_i'} \left(1 +\frac{1}{2}
S_{\alpha_i'\beta_i'}(R_i) \cdot \frac{\partial}{\partial
P_i}\right) \delta_{\alpha_i\beta_i}
\nonumber\\
&& - \frac{P_i}{M} \cdot d_{\alpha_i\beta_i}\left(1+\frac{1}{2}
S_{\alpha_i\beta_i}(R_i)\cdot  \frac{\partial}{\partial P_i}
\right) \delta_{\alpha_i'\beta_i'}\;,
\end{eqnarray}
where $S_{\alpha_i \beta_i}= (E_{\alpha_i}-E_{\beta_i})
d_{\alpha_i \beta_i} (\frac{P}{M}\cdot d_{\alpha_i \beta_i})^{-1}$
and $d_{\alpha_i \beta_i}=<\alpha_i;R|\nabla_R|\beta_i;R>$ is the
non-adiabatic coupling matrix element.

The formal solution of Eq.~(\ref{eq:eqofm}) is
\begin{equation}
W^{\alpha_1' \alpha_1 \alpha_2' \alpha_2}(t) =
\left(\exp({\frac{1}{2}\mbox{\boldmath ${\mathcal K}$} t}) {\bf
W}(0)\right)_{\alpha_1' \alpha_1 \alpha_2' \alpha_2}\;.
\label{eq:fsolneqofm}
\end{equation}
Now that the quantum-classical evolution equation for the matrix
elements of $W$ and its formal solution have been determined, we
can return to the calculation of the quantum-classical limit of
the quantum time correlation function.

\section{Quantum-Classical Correlation Function} \label{sec:qccor}

Equation~(\ref{eq:eqofm}) is one of the main results of this
paper. Using it we can obtain a quantum-classical approximation to
the quantum correlation function by replacing the full quantum
evolution of the spectral density in Eq.~(\ref{eq:corfin1}) with
its evolution in the quantum-classical limit given by
Eq.~(\ref{eq:fsolneqofm}), the solution of Eq.~(\ref{eq:eqofm}).
We have
\begin{eqnarray}
C_{AB}(t;\beta) &=& \frac{1}{\beta}\int_0^{\beta}d\lambda
\sum_{\alpha_1,\alpha_1',\alpha_2,\alpha_2'} \int \prod_{i=1}^2
dX_i \;  B_W^{\dag \alpha_1\alpha_1'}(X_1)
A_W^{\alpha_2\alpha_2'}(X_2)
\nonumber\\
&&\times \Big(\exp({\frac{1}{2}\mbox{\boldmath ${\mathcal K}$} t})
{\bf
W}(X_1,X_2,0;\beta,\lambda)\Big)_{\alpha_1'\alpha_1\alpha_2'\alpha_2}\;.
\label{eq:corfin}
\end{eqnarray}
Since the operator $\mbox{\boldmath ${\mathcal K}$}$ is the sum of
two operators, one acting only on functions of $X_1$ and quantum
indices with subscript 1, and the other on functions of $X_2$ and
quantum indices with subscript 2, we may integrate by parts to
have the operator act on the dynamical variables instead of ${\bf
W}$. We obtain,
\begin{eqnarray}
C_{AB}(t,\beta) &=& \sum_{\beta_1'\beta_1\beta_2'\beta_2} \int
\prod_{i=1}^2 dX_i \; B_W^{\dag \beta_1\beta_1'}(X_1,\frac{t}{2})
A_W^{\beta_2\beta_2'}(X_2,-\frac{t}{2})
\overline{W}^{\beta_1'\beta_1\beta_2'\beta_2}(X_1,X_2;\beta)\;,
\label{eq:qccor}
\end{eqnarray}
where
\begin{eqnarray}
B_W^{\dag \beta_1\beta_1'}(X_1,\frac{t}{2})&=&
\sum_{\alpha_1'\alpha_1} \Big(e^{i\frac{t}{2}{\cal
L}(X_1)}\Big)_{\beta_1\beta_1' \alpha_1\alpha_1' } \hat{B}_W^{\dag
\alpha_1\alpha_1'}(X_1)\;,
\nonumber\\
A_W^{\beta_2\beta_2'}(X_2,-\frac{t}{2})&=&
\sum_{\alpha_2'\alpha_2} \Big(e^{-i\frac{t}{2}{\cal
L}(X_2)}\Big)_{\beta_2\beta_2' \alpha_2\alpha_2' }
\hat{A}_W^{\alpha_2\alpha_2'}(X_2) \;.\label{eq:qcevol}
\end{eqnarray}
In writing Eq.~(\ref{eq:qccor}) we defined
\begin{equation}
\overline{W}^{\beta_1'\beta_1\beta_2'\beta_2}(X_1,X_2;\beta)=
\frac{1}{\beta}\int_0^{\beta} d \lambda \;
W^{\beta_1'\beta_1\beta_2'\beta_2}(X_1,X_2,0;\beta,\lambda)\;.
\label{eq:overW}
\end{equation}
Equation~(\ref{eq:qccor}) shows that the correlation function at
time $t$ can be calculated by sampling $X_1$ and $X_2$ from
suitable weights determined by
$\overline{W}^{\beta_1'\beta_1\beta_2'\beta_2}(X_1,X_2;\beta)$ at
time zero and propagating $B_W^{\dag \alpha_1\alpha_1'}$ forward
in time and $A_W^{\alpha_2\alpha_2'}$ backward in time for an
interval of length $t/2$. Note that while the time evolution in
Eq.~(\ref{eq:qccor}) is by quantum-classical dynamics, the initial
condition for
$\overline{W}^{\beta_1'\beta_1\beta_2'\beta_2}(X_1,X_2;\beta)$ is
still an exact expression for the full equilibrium quantum mechanical spectral
density.

\section{High Temperature Form of $\overline{\bf W}$ } \label{sec:icondt}

At $t=0$ ${\bf W}$ is given explicitly by
\begin{eqnarray}
W^{\alpha_1'\alpha_1\alpha_2'\alpha_2}(X_1,X_2,0;\beta,\lambda)
&=& \frac{1 }{(2\pi\hbar)^{2\nu_h}Z_Q} \int dZ_1dZ_2
e^{-\frac{i}{\hbar}(P_1Z_1+P_2Z_2)}
\nonumber\\
&\times&
\langle \alpha_1';R_1|\langle R_1+\frac{Z_1}{2}|
e^{-(\beta-\lambda)\hat{H}}
|R_2-\frac{Z_2}{2}\rangle|\alpha_2;R_2\rangle
\nonumber \\
&\times&
\langle \alpha_2';R_2|\langle R_2+\frac{Z_2}{2}|
e^{-\lambda\hat{H}}
|R_1-\frac{Z_1}{2}\rangle|\alpha_1;R_1\rangle \;.
\label{eq:iniW}
\end{eqnarray}
It can be computed using path integral techniques but
its evaluation is still a difficult problem.
In order to illustrate its structure we consider its form in the
high temperature limit.
In this limit we may write
\begin{eqnarray}
\langle R_2+\frac{Z_2}{2}|e^{-\lambda\hat{H}}|R_1-\frac{Z_1}{2}\rangle
&\approx&
e^{-\lambda\hat{h}(R_c-\frac{1}{4}Z_{12})}
\left(\frac{M}{2\pi\lambda\hbar^2}\right)^{\nu_h/2}
\exp\left[-\frac{M(R_{12}-Z_c)^2}{2\lambda\hbar^2}\right]
\;,\label{eq:ht-limit1}
\end{eqnarray}
where $\hat{h}=\frac{\hat{p}^2}{2m}+\hat{V}$
and we have introduced the variables $Z_c=(Z_1+Z_2)/2$, $Z_{12}=Z_1-Z_2$,
$R_c=(R_1+R_2)/2$ and $R_{12}=R_1-R_2$.
Taking the desired matrix element of this expression
and inserting complete sets of adiabatic states we obtain
\begin{eqnarray}
\langle\alpha_2';R_2|\langle R_2+\frac{Z_2}{2}|
e^{-\lambda\hat{h}}|R_1-\frac{Z_1}{2}\rangle|\alpha_1;R_1\rangle
&=&
\sum_{\alpha}e^{-\lambda E_{\alpha}(R_c-\frac{Z_{12}}{2})}
\langle\alpha_2';R_2|\alpha;R_c-\frac{Z_{12}}{4}\rangle
\langle\alpha;R_c-\frac{Z_{12}}{4}|\alpha_1;R_1\rangle \nonumber\\
&=&
\sum_{\alpha}e^{-\lambda E_{\alpha}(R_c)}
\langle\alpha_2';R_2|\alpha;R_c\rangle
\langle\alpha;R_c|\alpha_1;R_1\rangle +{\cal O}(Z_{12})\;.
\end{eqnarray}
Keeping only the zero order term in $Z_{12}$,
the integral over $Z_{12}$ in Eq.~(\ref{eq:iniW}) gives a factor
$(4\pi\hbar)^{\nu_h}\delta(P_1-P_2)$.
The other term  in Eq.~(\ref{eq:iniW}), which
arises from the combination of the gaussian on the right hand side
of Eq.~(\ref{eq:ht-limit1}) along with the analogous expression coming
from the high temperature limit of
$\langle R_1+\frac{Z_1}{2}|e^{-(\beta-\lambda)\hat{H}}
|R_2-\frac{Z_2}{2}\rangle$, can be evaluated by
performing the gaussian integral on $Z_c$ to obtain
\begin{eqnarray}
\left(\frac{M}{2\pi\hbar^2\beta}\right)^{\nu_h/2}
e^{\frac{i}{\hbar}2P_c
\frac{2\lambda-\beta}{\beta} R_{12}}
e^{-\frac{2M}{\hbar^2(\beta)}R_{12}^2}
e^{-\frac{\lambda(\beta-\lambda)}{\beta} \frac{2P_c^2}{M}}
&=&
\left(\frac{M}{2\pi\hbar^2\beta}\right)^{\nu_h/2}
f(R_{12},P_c)
e^{-\frac{\lambda(\beta-\lambda)}{\beta} \frac{2P_c^2}{M}}
\;,
\end{eqnarray}
where $P_c=(P_1+P_2)/2$. The function $f(R_{12},P_c)$ still contains
quantum information since it is composed of
a phase factor and a gaussian expressing quantum dispersion
effects in the heavy mass coordinates.
We can obtain a classical bath approximation if we represent $f(R_{12},P_c)$
in a multipole expansion and keep only the first order term,
$ f(R_{12},P_c)\approx[\int dR_{12} f(R_{12},P_c)]\delta(R_{12}) $,
we have
\begin{eqnarray}
f(R_{12},P_c)
&\approx&\left(\frac{\pi\hbar^2\beta}{2M}\right)^{\nu_h/2}
e^{-\frac{P_c^2}{2\beta M}(2\lambda-\beta)^2}\delta(R_{12})\;
\end{eqnarray}
Combining terms we obtain a high-temperature, classical-bath approximation to
$\bf W$:
\begin{eqnarray}
W^{\alpha_1'\alpha_1\alpha_2'\alpha_2}(X_1,X_2,0;\beta,\lambda)
&\approx& \frac{1}{(2\pi\hbar)^{\nu_h}Z_Q} e^{-\beta
\frac{P_1^2}{2M}} e^{-(\beta-\lambda)E_{\alpha_1'}(R_1)}
e^{-\lambda E_{\alpha_2'}(R_1)} \delta_{\alpha_1'\alpha_2}
\delta_{\alpha_2' \alpha;R_1} \delta(R_{12})\delta(P_{12})\;.
\end{eqnarray}
Thus the quantity $\overline{\bf W}$, defined in Eq.~(\ref{eq:overW}),
is given
in the high-temperature, classical-bath limit by
\begin{eqnarray}
\overline{W}^{\alpha_1'\alpha_1\alpha_2'\alpha_2}(X_1,X_2;\beta)
&=&
\frac{1 }{(2\pi\hbar)^{\nu_h}Z_Q}
e^{-\beta\left(\frac{P_1^2}{2M} + E_{\alpha_1'}(R_1)\right)}
\frac{e^{\beta(E_{\alpha_1'}(R_1)-E_{\alpha_2'}(R_1))}-1}
{\beta(E_{\alpha_1'}(R_1)-E_{\alpha_2'}(R_1))}
\nonumber\\
&\times&
\delta_{\alpha_1'\alpha_2} \delta_{\alpha_2' \alpha;R_1}
\delta(R_{12})\delta(P_{12})
\;.\label{eq:barWhight}
\end{eqnarray}
Using similar manipulations, the
high temperature limit of $Z_Q$ is
\begin{equation}
Z_Q \approx \frac{1}{(2\pi\hbar)^{\nu_h}}
\sum_{\alpha}\int dRdPe^{-\beta\left(\frac{P^2}{2M}+E_{\alpha}(R)\right)}\;.
\end{equation}
If Eq.~(\ref{eq:barWhight}) is used in the correlation function
formula, Eq.~(\ref{eq:qccor}), the result maybe shown
to correspond with the quantum-classical linear response
theory form \cite{mqcstat} to lowest order in $\hbar$.

\section{Conclusion} \label{sec:conc}
The expression for the quantum-classical limit of the quantum
correlation function derived in this paper provides a route for
the calculation of quantum transport properties in condensed phase
systems. Difficult many-body quantum dynamics is replaced by
quantum-classical evolution which can be carried out using
surface-hopping schemes involving probabilistic sampling of
quantum transitions, with associated momentum changes in the bath,
and classical trajectory segments. The classical trajectory
segments are accompanied by phase factors that account for quantum
coherence when off-diagonal matrix elements appear.
\cite{kapral:1} The full equilibrium quantum structure of the
entire system is retained. While the equilibrium calculation is
still a difficult problem it is more tractable than the quantum
dynamics needed to treat the many-body system using full quantum
dynamics. For example, imaginary time Feynman path integral
methods for computing equilibrium properties are far more
tractable than their corresponding real time variants. Since
quantum information about the entire system is retained in the
equilibrium structure, the formula for the correlation function
incorporates some aspects of nuclear bath quantum dispersion that
is missing in other quantum-classical schemes. The importance of
retaining the full quantum equilibrium structure has been noted in
Ref.~\cite{egorov2}.

The results also provide a framework for exploring and extending
the statistical mechanics of quantum-classical systems. The
correlation functions for transport properties that result from
linear response theory in quantum-classical systems involve both
quantum-classical evolution like that derived in this paper, as
well as the equilibrium quantum-classical density that is
stationary under the quantum-classical evolution.
\cite{mqcstat,simu} One may construct approximations to quantum
transport properties by considering other approximate limiting
forms of the equilibrium spectral density.
We also note that to establish a complete comparison
with quantum-classical linear response theory requires
the retention of terms that were neglected in the calculations for
$\overline{\bf W}$ presented in Sec.~\ref{sec:icondt}.
It should be fruitful to pursue extensions of such
calculations to obtain other approximations for quantum transport
properties.

\section*{Acknowledgements}
This  work  was  supported in part by a grant from the Natural
Sciences and Engineering Research Council of Canada. RK would like
to thank S. Bonella and G. Ciccotti for discussions pertaining to
part of the work presented here.

\appendix

\section{Derivation of Quantum-Classical Evolution Equation for $W$}
\label{app:qcW}

The equation of motion for the spectral density
(Eq.~(\ref{eq:partW})) takes a similar form in scaled coordinates:
\begin{eqnarray}
\frac{\partial }{\partial
t}W'(x_1',x_2',X_1',X_2',t';\beta',\lambda')&=&
-\frac{1}{2}\Big(iL_1^{(0)\prime}(x_1',X_1')-iL_2^{(0)\prime}(x_2',X_2')\Big)
W'(x_1',x_2',X_1',X_2',t';\beta',\lambda')
\nonumber \\
&+&\frac{1}{2}\int \prod_{i=1}^2 ds_i' dS_i' \;
\Big(\omega_1'(r_1',s_1',R_1',S_1')\delta(s_2')\delta(S_2')
- \omega_2'(r_2',s_2',R_2',S_2')\delta(s_1')\delta(S_1')\Big)
\nonumber \\
&\times&
W'(x_1'-\pi_1',X_1'-\Pi_1',x_2'-\pi_2',X_2'-\Pi_2',t';\beta',\lambda')
\;,\label{eq:scaledpartW}
\end{eqnarray}
where the scaled free streaming Liouville operator and integral
kernel are defined in Eqs.~(\ref{eq:scaledL0}) and
(\ref{eq:scaledomega}), respectively.

Inserting Eq.~(\ref{eq:potexp}) into the expression for
$\omega_1'$ and retaining only terms up to linear order in in
$\mu$ we find
\begin{eqnarray}
\omega_1'(r_1',s_1',R_1',S_1') &\approx& \frac{2}{(\pi)^{\nu}}\int
d\bar{r}_1'd\tilde{R}_1' V'(r_1'-\bar{r}_1',R_1') \sin\left(2s_1'
\cdot \bar{r}_1' +2S_1' \cdot \tilde{R}_1'\right)
\nonumber \\
&& - \frac{2\mu}{(\pi)^{\nu}}\int d\bar{r}_1'd\tilde{R}_1'
\frac{\partial V'(r_1'-\bar{r}_1',R_1')}{\partial R_1'} \cdot
\tilde{R}_1' \sin\left(2s_1' \cdot \bar{r}_1' +2S_1' \cdot
\tilde{R}_1'\right) +{\cal O}(\mu^2) \;.
\end{eqnarray}

We observe that
\begin{eqnarray}
\int d\bar{r}_1'd\tilde{R}_1' V'(r_1'-\bar{r}_1',R_1')
\sin\left(2s_1'\cdot \bar{r}_1' +2S_1'\cdot \tilde{R}_1'\right)
&=& \int d\bar{r}_1'd\tilde{R}_1' V'(r_1'-\bar{r}_1',R_1') \left[
\sin(2s_1'\cdot \bar{r}_1')\cos(2S_1'\cdot \tilde{R}_1')
\right. \nonumber \\
&&\left.  +\cos(2s_1'\cdot \bar{r}_1')\sin(2S_1'\cdot
\tilde{R}_1') \right]\;,
\end{eqnarray}
using the trigonometric identity for the sine of a sum of
arguments. Then, using the fact that $\int
d\tilde{R}_1'cos(2S_1'\tilde{R}_1')=\pi^{\nu_{\rm h}}\delta(S_1')$
we have
\begin{eqnarray}
\int d\bar{r}_1'd\tilde{R}_1' V'(r_1'-\bar{r}_1',R_1')
\sin\left(2s_1' \cdot \bar{r}_1'+2S_1' \cdot \tilde{R}_1'\right)
&=&\pi^{\nu_{\rm h}}\int d\bar{r}_1'V'(r_1'-\bar{r}_1',R_1')
\sin(2s_1' \cdot \bar{r}_1')\delta(S_1')\;,
\end{eqnarray}
where we have used $\int d\tilde{R}_1'\sin(2S_1'\cdot
\tilde{R}_1')=0$. In a similar manner
\begin{eqnarray}
\int d\bar{r}_1'd\tilde{R}_1' \frac{\partial
V'(r_1'-\bar{r}_1',R_1')}{\partial R_1'} \cdot \tilde{R}_1'
\sin\left(2s_1' \cdot \bar{r}_1'+2S_1' \cdot \tilde{R}_1'\right)
&=& -\frac{\pi^{\nu_{\rm h}}}{2}\int d\bar{r}_1' \frac{\partial
V'(r_1'-\bar{r}_1',R_1')}{\partial R_1'} \cos(2s_1' \cdot
\bar{r}_1') \cdot \frac{d\delta(S_1')}{dS_1'} \;,\nonumber \\
\end{eqnarray}
where we have used the relations $\int
d\tilde{R}_1'\tilde{R}_1'\cos(2S_1'\cdot \tilde{R}_1')=0$ and
$\int d\tilde{R}_1'\tilde{R}_1'\sin(2S_1'\cdot \tilde{R}_1')=
-(\pi^{\nu_{\rm h}}/2)d\delta(S_1')/dS_1'$. Then to order ${\cal
O}(\mu)$,
\begin{eqnarray}
\omega_1'(r_1',s_1',R_1',S_1') &=& \frac{2}{\pi^{\nu_{\ell}}}\int
d\bar{q}'V'(r_1'-\bar{r}_1',R_1')\sin(2s_1' \cdot \bar{r}_1')\delta(S_1')\nonumber\\
&+&\mu\left[ \frac{1}{\pi^{\nu_{\ell}}} \int
d\bar{r}_1'\frac{\partial V'(r_1'-\bar{r}_1',R_1')}{\partial R_1'}
\cos(2s_1' \cdot \bar{r}_1') \cdot
\frac{d\delta(S_1')}{dS_1'}\right] \;. \label{A:omega_1'}
\end{eqnarray}

Using this expression we may compute the integral on the right
hand side of Eq.~(\ref{eq:scaledpartW}) involving $\omega_1'$. The
algebra for the $\omega_2'$ term is similar. Given the
expression~(\ref{A:omega_1'}), the integral
\begin{eqnarray}
\int ds_1'dS_1'\omega_1'(r_1',s_1',R_1',S_1')
W'(x_1'-\pi_1',X_1'-\Pi_1',x_2',X_2',t';\beta,\lambda)\;,
\end{eqnarray}
has two contributions. The first is is
\begin{eqnarray}
&&\frac{2}{\pi^{\nu_{\ell}}} \int
ds_1'dS_1'W'(x_1'-\pi_1',X_1'-\Pi_1',x_2',X_2',t';\beta,\lambda)
\int d\bar{r}_1'V'(r_1'-\bar{r}_1',R_1')\sin(2s_1' \cdot
\bar{r}_1')\delta(S_1') = \nonumber \\
&&\qquad \frac{2}{\pi^{\nu_{\ell}}} \int
ds_1'W'(x_1'-\pi_1',X_1',x_2',X_2';t',\beta,\lambda) \int
d\bar{r}_1'V'(r_1'-\bar{r}_1',R_1')\sin(2s_1' \cdot \bar{r}_1')
\;,
\end{eqnarray}
while the second is
\begin{eqnarray}
\frac{\mu}{\pi^{\nu_{\ell}}} \int
ds_1'dS_1'W'(x_1-\pi_1',X_1'-\Pi_1',x_2',X_2',t';\beta,\lambda)
\int d\bar{r}_1'\frac{\partial }{\partial
R_1'}V'(r_1'-\bar{r}_1',R_1') \cos(2s_1' \cdot \bar{r}_1') \cdot
\frac{d\delta(S_1')}{dS_1'}
&=&\nonumber \\
\frac{\mu}{\pi^{\nu_{\ell}}} \int ds_1'\frac{\partial} {\partial
P_1'} W'(x_1'-\pi_1',X_1',x_2',X_2',t';\beta,\lambda) \cdot \int
d\bar{r}_1'  \frac{\partial }{\partial
R_1'}V'(r_1'-\bar{r}_1',R_1') \cos(2s_1' \cdot \bar{r}_1')  \;.
\end{eqnarray}
Defining $\Delta F'_{R_1',s_1'}=-\frac{\partial}{\partial R_1'}
\left[V'(R_1')\delta(s_1') -\frac{1}{\pi^{\nu_{\ell}}} \int
d\bar{r}_1'\cos(2s_1'\cdot \bar{r}_1')V'(r_1'-\bar{r}_1',R_1')
\right]$ and returning to unscaled coordinates we obtain
Eq.~(\ref{eq:qcwignerW}), the desired quantum-classical evolution
equation for the full Wigner representation of $W$. The first term
in the definition of $\Delta F'$ is compensated by the
introduction of the full classical propagator for the heavy mass
degrees of freedom. Written in this form, $\Delta F'$ also depends
only on the interaction potential between the light and heavy mass
particles.

\section{Equation in the partial Wigner representation}
\label{t-1utw}

In this appendix we perform explicitly the calculations
that, starting from Eq.~(\ref{eq:qcwignerW}),
lead to Eq.~(\ref{eq:eqofm}).
This calculation amounts to the evaluation of
$({\cal T}^{-1}\circ K \circ{\cal T})\circ {\bf W}(t)$.
The various terms in $K$, defined by Eq.~(\ref{eq:qcwignerW}),
are considered separately.

Consider  the calculation of
$({\cal T}^{-1}\circ i L_1(X_1)\circ{\cal T})\circ {\bf W}$
which is composed of two terms. The force term is
${\cal T}^{-1}\circ F_{R_1}\cdot\partial/\partial P_1{\cal T}\circ {\bf W}=
F_{R_1}\cdot\partial/\partial P_1{\bf W}$,
since ${\cal T}$ and its inverse do not depend on $P_1$.
The free streaming term $({\cal T}^{-1}\circ \frac{P_1}{M}\cdot
\frac{\partial }{\partial R_1}\circ {\cal T}) \circ{\bf W}$
requires additional calculations since ${\cal T}$ depends on $R_1$.
We have
\begin{eqnarray}
({\cal T}^{-1}\circ\frac{P_1}{M}\cdot\frac{\partial }{\partial R_1}&\circ&
{\cal T})\circ {\bf W}
=
\frac{P_1}{M}\cdot
\sum_{\beta_1\beta_1'\beta_2\beta_2'}
\int\prod_{j=1}^2 dr_jdz_j
\phi_{\alpha_1'}^*(r_1+\frac{z_1}{2},R_1)\phi_{\alpha_1}(r_1-\frac{z_1}{2},R_1)
\nonumber \\
&\times&
\phi_{\alpha_2'}^*(q_2+\frac{z_2}{2};R_2)\phi_{\alpha_2}(r_2-\frac{z_2}{2};R_2)
\phi_{\beta_2'}(r_2+\frac{z_2}{2};R_2)\phi_{\beta_2}^*(q_2-\frac{z_2}{2};R_2)
\nonumber\\
&\times&
\left\{\left[
\frac{\partial \phi_{\beta_1'}(r_1+\frac{z_1}{2};R_1)}{\partial R_1}
\phi_{\beta_1}^*(r_1-\frac{z_1}{2};R_1)+\phi_{\beta_1'}(r_1+\frac{z_1}{2};R_1)
\frac{\partial \phi_{\beta_1}^*(r_1-\frac{z_1}{2};R_1)}{\partial R_1}
\right] W^{\beta_1'\beta_1\beta_2'\beta_2}\right.
\nonumber\\
&+&\left.\phi_{\beta_1'}(r_1+\frac{z_1}{2};R_1)
\phi_{\beta_1}^*(r_1-\frac{z_1}{2};R_1)
\frac{\partial}{\partial R_1}W^{\beta_1'\beta_1\beta_2'\beta_2}\right\}
\;.\label{eq:appb1}
\end{eqnarray}
The last term where $\partial{\bf W}/\partial R_1$ appears
is simple to calculate and gives
$\frac{P_1}{M}\cdot\frac{\partial}{\partial R_1}
W^{\alpha_1'\alpha_1\alpha_2'\alpha_2}$.
To calculate the other two terms, we make the change
of variables
$q_1=r_1-\frac{z_1}{2}$, $ q_2=r_1+\frac{z_1}{2}$,
$q_3=r_2-\frac{z_2}{2}$, $q_4=r_2+\frac{z_2}{2}$.
Integrating over $q_3$ and $q_4$ and using
$\int dq\phi_{\alpha}(q,R)\phi_{\beta}^*(q,R)=\delta_{\alpha\beta}$,
we obtain
\begin{eqnarray}
&&
\frac{P_1}{M}\cdot
\sum_{\beta_1\beta_1'}
\int dq_1dq_2 \phi_{\alpha_1'}^*(q_2;R_1)\phi_{\alpha_1}(q_1;R_1)
\left[\frac{\partial \phi_{\beta_1'}(q_2;R_1)}{\partial R_1}
\phi_{\beta_1}^*(q_1;R_1)+\phi_{\beta_1'}(q_2;R_1)
\frac{\partial \phi_{\beta_1}^*(q_1;R_1)}{\partial R_1}
\right] W^{\beta_1'\beta_1\alpha_2'\alpha_2}
\nonumber\\
&&= \frac{P_1}{M}\cdot
\sum_{\beta_1'} d_{\alpha_1'\beta_1'} W^{\beta_1'\alpha_1\alpha_2'\alpha_2}
- \frac{P_1}{M}\cdot \sum_{\beta_1} d_{\beta_1\alpha_1}
W^{\alpha_1'\beta_1\alpha_2'\alpha_2}\;,
\end{eqnarray}
where we have introduced the definition of the nonadiabatic coupling vector.

Consider the calculation of $({\cal T}^{-1}\circ iL_1^{(0)}(x_1)
\circ{\cal T})\circ {\bf W}$. In this case one integration over the
momentum, arising from the definition of ${\cal T}^{-1}$,
gives
$
\int dp_1(p_1/m)e^{\frac{i}{\hbar}p_1(z_1-z_3)}
=(2\pi\hbar)^{\nu_{\ell}}(i\hbar/m)
\partial\delta(z_1-z_3)/\partial z_3 $,
while integration over $p_2$ gives $(2\pi\hbar)^{\nu_{\ell}}\delta(z_2-z_4)$.
One can then integrate by parts on $z_3$ and obtain
\begin{eqnarray}
({\cal T}^{-1}\circ\frac{p_1}{m}&&\frac{\partial }{\partial r_1}
\circ{\cal T})\circ {\bf W}
=-\frac{i\hbar}{m}\sum_{\beta_1\beta_1'\beta_2\beta_2'}
\int \prod_{j=1}^2dr_j\prod_{k=1}^3dz_k
\delta(z_1-z_3)
\phi_{\alpha_1'}^*(r_1+\frac{z_1}{2};R_1)\phi_{\alpha_1}(r_1-\frac{z_1}{2};R_1)
\nonumber\\ &&\times
\phi_{\alpha_2'}^*(r_2+\frac{z_2}{2};R_2)\phi_{\alpha_2}(r_2-\frac{z_2}{2};R_2)
\phi_{\beta_2'}(r_2+\frac{z_2}{2};R_2)\phi_{\beta_2}^*(r_2-\frac{z_2}{2};R_2)
\nonumber\\
&&\times
\frac{\partial^2}{\partial z_3\partial r_1}
\left[
\phi_{\beta_1'}(r_1+\frac{z_3}{2};R_1)\phi_{\beta_1}^*(r_1-\frac{z_3}{2};R_1)
\right] W^{\beta_1'\beta_1\beta_2'\beta_2}\;.
\end{eqnarray}
Then use
\begin{eqnarray}
\frac{\partial^2}{\partial z_3\partial r_1}
&&\left[ \phi_{\beta_1'}(r_1+\frac{z_3}{2};R_1)
\phi_{\beta_1}^*(r_1-\frac{z_3}{2};R_1)
\right] \nonumber\\
&&= \frac{1}{2} \left[
\frac{\partial^2\phi_{\beta_1'}(r_1+\frac{z_3}{2};R_1)}
{\partial(r_1+\frac{z_3}{2})^2}\phi_{\beta_1}^*(r_1-\frac{z_3}{2};R_1)
-\phi_{\beta_1'}(r_1+\frac{z_3}{2};R_1)
\frac{\partial^2\phi_{\beta_1}^*(r_1-\frac{z_3}{2};R_1)}
{\partial(r_1-\frac{z_3}{2})^2} \right]\;,
\end{eqnarray}
go back to the integral, perform the integration on $z_3$
and make the change of variables previously introduced.
The integration over $q_3$ and $q_4$ can be performed
using the completeness of the adiabatic basis and one gets
\begin{eqnarray}
({\cal T}^{-1}\circ\frac{p_1}{m}&&\frac{\partial }{\partial r_1}
\circ{\cal T})\circ{\bf W}
=\frac{i}{\hbar}\sum_{\beta_1'}\langle\alpha_1';R_1|\frac{\hat{p}^2}{2m}
|\beta_1';R_1\rangle
W^{\beta_1'\alpha_1\alpha_2'\alpha_2}
\nonumber\\
&&
-\frac{i}{\hbar}\sum_{\beta_1}\langle\beta_1;R_1|\frac{\hat{p}^2}{2m}
|\alpha_1;R_1\rangle
W^{\alpha_1'\beta_1\alpha_2'\alpha_2}\;,
\end{eqnarray}
where we have used the identity
$
\langle\alpha,R|\hat{p}^2/2m|\beta,R\rangle
=-(\hbar^2/2m)\int dq\phi_{\alpha}^*(q,R)
\partial\phi_{\beta}(q,R)/\partial q^2
$.

We consider now the transformation of the potential term
in $K\circ W$ equal to
$c_1\int ds_1ds_2\delta(s_2)\int d\bar{r}
V(r_1-\bar{r})\sin\left(\frac{2s_1\cdot\bar{r}}{\hbar}\right)
W(x_1-\pi_1,x_2-\pi_2,X_1,X_2,t;\beta\lambda)$, where
$c_1=2\hbar^{-1}(\pi\hbar)^{-\nu_{\ell}}$;
we may immediately perform the trivial integrations on $p_1$, $p_2$,
$z_3$ and $z_4$. We also make the change variables
below Eq.~(\ref{eq:appb1})
so that the integrations on
$q_3$ and $q_4$ are also trivial. One is left with the integral
\begin{eqnarray}
&& c_1 \sum_{\beta_1'\beta_1} \int dq_1dq_2
\int ds_1d\bar{r}V(\frac{q_1+q_2}{2}-\bar{r})
\sin\left(\frac{2s_1\cdot\bar{r}}{\hbar}\right)
e^{\frac{i}{\hbar}s_1\cdot(q_2-q_1)}
\nonumber\\
&&\times
\phi_{\alpha_1'}^*(q_2;R_1)\phi_{\alpha_1}(q_1;R_1)
\phi_{\beta_1'}(q_2;R_1)\phi_{\beta_1}^*(q_1;R_1)
W^{\beta_1'\beta_1\alpha_2'\alpha_2}
\;. \label{eq:integral_potential}
\end{eqnarray}
Using the fact that
$ \int ds_1 e^{\frac{i}{\hbar}s_1\cdot(q_2-q_1)}
\sin\left(2s_1\cdot\bar{r}/\hbar\right)
=(2\pi\hbar)^{\nu_{\ell}} \left[\delta(q_2-q_1+2\bar{r})-
\delta(q_2-q_1-2\bar{r})\right]/2i $,
substituting into Eq.~(\ref{eq:integral_potential})
and making the change of variable $\sigma=2\bar{r}$,
the delta functions can be integrated out and one obtains,
\begin{eqnarray}
&&
\frac{1}{i\hbar}
\sum_{\beta_1'\beta_1}
\left[\int dq_1 \phi_{\alpha_1}(q_1;R_1) \phi_{\beta_1}^*(q_1;R_1) \right]
\left[
\int dq_2\phi_{\alpha_1'}^*(q_2;R_1) V(q_2) \phi_{\beta_1'}(q_2;R_1)\right]
W^{\beta_1'\beta_1\alpha_2'\alpha_2}
\nonumber\\
&&-
\frac{1}{i\hbar}
\sum_{\beta_1'\beta_1}
\left[\int dq_1
\phi_{\beta_1}^*(q_1;R_1)V(q_1)\phi_{\alpha_1}(q_1;R_1)\right]
\left[\int dq_2\phi_{\alpha_1'}^*(q_2;R_1)
\phi_{\beta_1'}(q_2;R_1)\right]
W^{\beta_1'\beta_1\alpha_2'\alpha_2}
\nonumber\\
&& =
\frac{1}{i\hbar}
\sum_{\beta_1'}
\langle \alpha_1';R_1| \hat{V}|\beta_1';R_1\rangle
W^{\beta_1'\alpha_1\alpha_2'\alpha_2}
- \frac{1}{i\hbar}
\sum_{\beta_1}
\langle \beta_1;R_1|\hat{V}|\alpha_1;R_1\rangle
W^{\alpha_1'\beta_1\alpha_2'\alpha_2}
\;.
\end{eqnarray}

The last term that must be worked out explicitly
arises from the transformation of the quantum-classical term
$\int ds_1ds_2\delta(s_2)
\Delta F_1(R_1,s_1)\cdot\frac{\partial}{\partial P_1}
W(x_1-\pi_1,x_2-\pi_2,X_1,X_2,t;\lambda,\beta)$.
Recalling the expression for $\Delta F_1$ in Eq.~(\ref{eq:DeltaF_i})
one sees that there are two contributions to transform.
The transformation of the
term involving $\frac{\partial V(R_1)}{\partial R_1}\delta(s_1)$
is the same as that for the force term in $iL_1(X_1)$ and yields
$F_{R_1}\cdot\partial W^{\alpha_1'\alpha_1\alpha_2'\alpha_2}/
\partial P_1$.
The integral term in $\Delta F_1(R_1,s_1)$ can be computed
by integrating over $p_1$, $p_2$, $z_3$ and $z_4$,
performing the change of variables
below Eq.~(\ref{eq:appb1})
and integrating over $q_3$ and $q_4$, to obtain
\begin{eqnarray}
&&
\sum_{\beta_1'\beta_1}
\frac{1}{(\pi\hbar)^{\nu_{\ell}}}\int dq_1dq_2
\phi_{\alpha_1'}^*(q_2;R_1)\phi_{\alpha_1}(q_1;R_1)
\phi_{\beta_1'}(q_2;R_1)\phi_{\beta_1}^*(q_1;R_1)
\nonumber \\
&&\times
\int ds_1\int d\bar{r}
\left(\frac{\partial }{\partial R_1}V(\frac{q_1+q_2}{2}-\bar{r},R_1)\right)
\cos\left(\frac{2s_1\cdot\bar{r}}{\hbar}\right)
e^{\frac{i}{\hbar}s_1\cdot(q_2-q_1)}
\frac{\partial}{\partial P_1}W^{\beta_1'\beta_1\alpha_2'\alpha_2} \;.
\end{eqnarray}
Then one can use the integral
$\int ds_1\cos\left(2s_1\cdot\bar{r}/\hbar\right)
e^{\frac{i}{\hbar}s_1\cdot(q_2-q_1)}
= (2\pi\hbar)^{\nu_{\ell}}
\left[ \delta(q_2-q_1+2\bar{r})+ \delta(q_2-q_1-2\bar{r})\right]/2$,
make the change of variable $\sigma=2\bar{r}$, and integrate
out the delta functions to find
\begin{equation}
\frac{1}{2}
\left[ \sum_{\beta_1'}
\langle\alpha_1';R_1| \frac{\partial\hat{V}(R_1)}{\partial R_1} |\beta_1';R_1\rangle
\frac{\partial}{\partial P_1}W^{\beta_1'\alpha_1\alpha_2'\alpha_2}
+
\sum_{\beta_1}
\langle\beta_1;R_1| \frac{\partial\hat{V}(R_1)}{\partial R_1}| \alpha_1;R_1\rangle
\frac{\partial}{\partial P_1}W^{\alpha_1'\beta_1\alpha_2'\alpha_2} \right]
\;.
\end{equation}

Analogous terms (but with opposite sign) are obtained
when considering the transformation of the terms depending
on $x_2$ and $X_2$ in Eq.~(\ref{eq:qcwignerW}).

Combining all terms and using the relations
$F_W^{\alpha\beta}(R)=-\langle\alpha;R|\nabla_R\hat{V}(R)|\beta;R\rangle
=F_W^{\alpha}+(E_{\alpha}-E_{\beta})d_{\alpha\beta}$,
\begin{eqnarray}
-\frac{i}{\hbar}\sum_{\beta_1'}\langle\alpha_1';R_1|
\left(\frac{\hat{p}^2}{2m}+\hat{V}\right)
|\beta_1';R_1\rangle W^{\beta_1'\alpha_1\alpha_2'\alpha_2}
&&+
\frac{i}{\hbar}\sum_{\beta_1}\langle\beta_1;R_1|
\left(\frac{\hat{p}^2}{2m}+\hat{V}\right)
|\alpha_1;R_1\rangle
W^{\alpha_1'\beta_1\alpha_2'\alpha_2}
=\nonumber\\
=-
\frac{i}{\hbar}(E_{\alpha_1'}(R_1)-E_{\alpha_1}(R_1))
W^{\alpha_1'\alpha_1\alpha_2'\alpha_2}
&&=-i\omega_{\alpha_1'\alpha_1}(R_1)
W^{\alpha_1'\alpha_1\alpha_2'\alpha_2}\;,
\end{eqnarray}
and introducing the definition
$S_{\alpha_i\beta_i}=(E_{\alpha_i}-E_{\beta_i})d_{\alpha_i\beta_i}
(\frac{P}{M}\cdot d_{\alpha_i\beta_i})^{-1}$, we find
Eq.~(\ref{eq:eqofm}).



\end{document}